\documentclass[12pt]{article}

\usepackage{amsmath}
\usepackage{amsfonts}
\usepackage{amssymb}
\usepackage{latexsym}
\usepackage{wasysym}
\usepackage{appendix}
\usepackage{setspace}
\usepackage{graphicx}		
\usepackage{subfigure}
\usepackage{epsfig}
\usepackage{rotating}

\usepackage{axodraw}

\begin{document}

\begin{flushright}
MAN/HEP/2012/06\\
July 2012
\end{flushright}

\bigskip\bigskip

\begin{center}
{\bf \Large{Anomalous Fermion Mass Generation at Three Loops} }
\end{center}

\bigskip\bigskip

\begin{center}
{\large Apostolos Pilaftsis}~\footnote[1]{E-mail address: {\tt
apostolos.pilaftsis@manchester.ac.uk}}\\[3mm] 
{\it Consortium for Fundamental Physics, School of Physics
and Astronomy,\\ University of Manchester, Manchester M13 9PL, 
United Kingdom}
\end{center}

\bigskip\bigskip\bigskip\bigskip

\centerline{\bf ABSTRACT} 

\noindent
We  present a novel  mechanism for  generating fermion  masses through
global anomalies at  the three loop level.  In  a gauge theory, global
anomalies  are  triggered  by  the  possible existence  of  scalar  or
pseudoscalar  states   and  heavy  fermions,  whose   masses  may  not
necessarily   result   from   spontaneous  symmetry   breaking.    The
implications of this mass-generating  mechanism for model building are
discussed,  including the  possibility of  creating  low-scale fermion
masses by quantum gravity effects.

\medskip
\noindent
{\small {\sc Keywords:} fermion mass generation; global anomalies;
  higher order loop effects}

\thispagestyle{empty}

\newpage

Anomalies represent  a profound  phenomenon in quantum  physics, where
symmetries  that may  occur in  the classical  action are  violated by
quantum-mechanical effects. The presence of anomalous terms in quantum
field        theory        was        originally       noted        by
Steinberger~\cite{Steinberger:1949wx}      and     subsequently     by
Schwinger~\cite{Schwinger:1951xk}.         However,       it       was
Adler~\cite{Adler:1969gk}, and  Bell and Jackiw~\cite{Bell:1969ts} who
first  unravelled the  remarkable properties  of the  so-called chiral
gauge  anomalies and  showed their  prominent role  in  explaining the
observed  sizeable decay  rate of  $\pi^0 \to  \gamma\gamma$.  Besides
chiral  anomalies,   a~number  of  other   equally  important  quantum
anomalies   has  been   discovered   since  then,   such  as   scaling
anomalies~\cite{Ellis:1975ap,Shifman:1978zn},             gravitational
anomalies~\cite{AlvarezGaume:1983ig} and the mixed gauge-gravitational
anomalies~\cite{Bardeen:1984pm,AlvarezGaume:1984dr}.

In this  paper we present  a novel radiative mechanism  for generating
fermion  masses at  the three  loop  level, by  means of  perturbative
chiral global anomalies.  In the context of a gauge theory, the chiral
global  anomalies are mediated  by scalar  or pseudoscalar  states and
heavy fermions, which may possess gauge-invariant bilinear masses that
do not originate from  a spontaneous symmetry breaking mechanism.  For
illustration,  we will  first analyze  an effective  scenario  for our
three-loop mass-generating mechanism, in  the sense that the generated
fermion  masses are  logarithmically ultra-violet  (UV)  divergent and
would  require a  UV cut-off  $\Lambda$.  The  cut-off scale~$\Lambda$
could  be  taken  to  be  the  Planck mass  $M_{\rm  Pl}  =  1.9\times
10^{19}$~GeV.   After  having   calculated  the  three-loop  radiative
fermion  mass within  the  framework  of an  effective  theory with  a
cut-off~$\Lambda$,  we will then  discuss a  minimal UV  completion of
this   effective  scenario.    Finally,  we   will   present  possible
applications   of  the   three-loop   mass-generating  mechanism   for
model-building.

To  start with,  we  first show  how  the anomalous  generation of  an
effective fermion  mass arises in a  simple model that can  serve as a
prototype  for  our discussion.   For  this  purpose,  we consider  an
anomalous-free U(1)$_V$ gauge theory with a singlet pseudoscalar state
$a$ and  two Dirac  fermions~$f$ and ~$F$,  which both couple  to $a$.
The fermion $F$ has a large U(1)$_V$-invariant mass $m_F$, whereas the
other fermion~$f$  is assumed to be  massless at the  tree level.  The
relevant Lagrangian of such a theory is given by
\begin{equation}
  \label{eq:LU1}
{\cal L}_{\rm } \ =\ -\,\frac{1}{4}\, F_{\mu\nu}\,F^{\mu\nu}\: +\: 
\bar{F}\, \Big(i\!\not\!\! D - m_F\Big) F\: +\: \bar{f}\, i\!\not\!\!D f\: 
+\: \frac{1}{2}\,(\partial_\mu a)(\partial^\mu a)\: +\:
a\,\Big( h_F\,\bar{F} i\gamma_5 F\: +\: h_f\,\bar{f} i\gamma_5 f\:\Big)\; ,
\end{equation}
where $F_{\mu\nu}  = \partial_\mu V_\nu  - \partial_\nu V_\mu$  is the
U(1)$_V$  field strength  tensor and  $D_\mu  = \partial_\mu  + i  g_V
Q_{F,f} V_\mu$ is the covariant  derivative acting on the fermions $F$
and $f$  with hypercharges $Q_F$ and $Q_f$,  respectively. In addition
to CP symmetry, Lagrangian~(\ref{eq:LU1}) has one additional symmetry,
in the absence of the fermion mass term $m_F \bar{F} F$. If $m_F = 0$,
the Lagrangian is invariant under the discrete transformations: $a \to
-a$,  $F_{R\,(L)} \to  +(-)\, F_{R\,(L)}$  and $f_{R\,(L)}  \to +(-)\,
f_{R\,(L)}$.  Since we consider $m_F  \neq 0$, this latter symmetry is
broken  softly   by  the   dimension-3  mass  operator   $m_F  \bar{F}
F$.  Finally, it  is important  to  remark that  the Yukawa  couplings
$h_{F,f}$ in the  Lagrangian~(\ref{eq:LU1}) break explicitly the shift
symmetry: $a \to a + c$, where $c$ is an arbitrary constant.

Our  aim is  now to  calculate the  three-loop effective  mass  of the
fermion~$f$  generated  by  virtue  of the  anomalous  operator:  $a\,
F^{\mu\nu}     \widetilde{F}_{\mu\nu}    \equiv     a\,    \frac{1}{2}
\varepsilon_{\mu\nu\lambda\rho}\,  F^{\mu\nu} F^{\lambda\rho}$, within
a UV-cutoff effective  theory.  In other words, we  will show that the
first  non-trivial  mixing  between  the  dimension-3  mass  operators
$\bar{F}  F$ and $\bar{f}  f$ occurs  at three  loops.  Then,  we will
present a  minimal UV  complete extension of  our simple  model.  Even
though this  is not essential for our  demonstration, the pseudoscalar
state  $a$ may also  have a  gauge-invariant bare  mass and  the gauge
field $V_\mu$ may become massive  through the usual realization of the
Higgs mechanism.

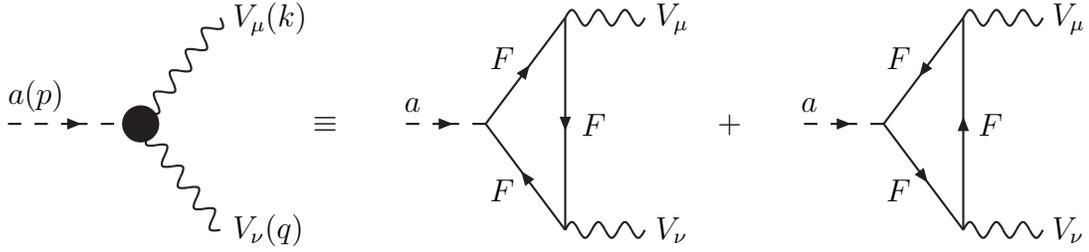
\begin{figure}
 
\begin{center}
\begin{picture}(400,100)(0,0)
\SetWidth{0.8}
 
\DashArrowLine(0,50)(50,50){5} \Vertex(50,50){7}
\Text(0,55)[bl]{$a (p)$}

\Photon(50,50)(80,90){3}{5}\Text(85,90)[l]{$V_\mu (k)$}
\Photon(50,50)(80,10){3}{5}\Text(85,10)[l]{$V_\nu (q)$}

\Text(120,50)[]{$\equiv$}

\DashArrowLine(150,50)(180,50){5} \Text(150,55)[bl]{$a$}

\ArrowLine(180,50)(210,90)\ArrowLine(210,90)(210,10)\ArrowLine(210,10)(180,50)
\Text(187,75)[]{$F$}\Text(187,25)[]{$F$}\Text(217,50)[l]{$F$}

\Photon(210,90)(240,90){3}{3} \Text(245,90)[l]{$V_\mu $}
\Photon(210,10)(240,10){3}{3} \Text(245,10)[l]{$V_\nu $}

\Text(273,50)[]{$+$}

\DashArrowLine(300,50)(330,50){5} \Text(300,55)[bl]{$a$}

\ArrowLine(360,90)(330,50)\ArrowLine(360,10)(360,90)\ArrowLine(330,50)(360,10)
\Text(337,75)[]{$F$}\Text(337,25)[]{$F$}\Text(367,50)[l]{$F$}

\Photon(360,90)(390,90){3}{3} \Text(395,90)[l]{$V_\mu$}
\Photon(360,10)(390,10){3}{3} \Text(395,10)[l]{$V_\nu$}

\end{picture}
\end{center}
\caption{\it The five-dimensional operator $a
  F_{\mu\nu}\widetilde{F}^{\mu\nu}$ induced by the chiral global
  anomaly of the heavy fermion $F$, with the convention $p + k + q =
  0$.}\label{fig:chiral}
\end{figure}

In the above U(1)$_V$ gauge theory, the pseudoscalar field $a$ couples
to  the  gauge  boson~$V_\mu$  via the  five-dimensional  operator  $a
F_{\mu\nu}  \widetilde{F}^{\mu\nu}$. This operator  is induced  by the
chiral global anomaly  of the heavy fermion $F$,  through the triangle
graphs shown  in Fig.~\ref{fig:chiral}.  With the  convention that all
momenta  are   incoming,  i.e.~$p  +  k   +  q  =   0$,  the  one-loop
$a(p)$-$V^\mu(k)$-$V^\nu(q)$ coupling
reads~\cite{Steinberger:1949wx,Adler:1969gk,Bell:1969ts}:
\begin{equation}
  \label{eq:aVV}
i\Gamma^{aVV}_{\mu\nu} (p,k,q)\ =\ i\, Q^2_F\, \frac{\alpha_V}{\pi}\; 
\frac{h_F}{m_F}\; F_P\bigg(\frac{p^2}{4m^2_F}\bigg)\;
\varepsilon_{\mu\nu\lambda\rho}\, k^\lambda q^\rho\; ,
\end{equation}
where $\varepsilon_{\mu\nu\lambda\rho}$ is the standard anti-symmetric
Levi--Civita  tensor  (with $\varepsilon^{0123}  =  +1$), $\alpha_V  =
g^2_V/(4\pi)$ is the fine structure constant associated with the gauge
field  $V_\mu$ and  $Q_F$ is  the  U(1)$_V$ hypercharge  of the  heavy
fermion~$F$.  Moreover, the loop function $F_P (\tau )$ may be written
down in one of the following forms~\cite{Steinberger:1949wx}:
\begin{eqnarray}
  \label{eq:FPtau}
F_P (\tau ) \ =\  
\int\limits_0^1 dx \int\limits_0^{1-x} dy\: \frac{2}{1\: -\: 
                     4(\tau +i\epsilon)\, x y}
\!& = &\! -\, \frac{1}{2(\tau + i\epsilon)}\,
\int\limits_0^1\frac{dx}{x}\: \ln \Big[\, 1\: -\: 
                     4(\tau +i\epsilon)\, x(1-x)\, \Big]\nonumber\\[3mm]
\!&&\! \hspace{-1.5cm}=\
\left\{ \begin{array}{lr}
\frac{\displaystyle 1}{\displaystyle \tau}\; \arcsin^2 \sqrt{\tau}\; ;
& |\tau| \leq 1\;,\\[2mm] 
-\frac{\displaystyle 1}{\displaystyle 4 \tau}\; \bigg[\, 
\ln\bigg(\frac{\displaystyle \sqrt{\tau} +
    \sqrt{\tau -1}}{\displaystyle \sqrt{\tau} -\sqrt{\tau -1}}\bigg) -
  i\pi\,\bigg]^2\; ; & |\tau| \geq 1\; ,  \end{array} \right.\qquad
\end{eqnarray}
with $\epsilon  \to 0^+$.   It is interesting  to quote the  small and
large argument limits of  the loop function~$F_P(\tau )$.  If~$|\tau |
\ll 1$, we have $F_P(\tau ) = 1 + \tau /3 + {\cal O}(\tau^2)$, whereas
in the opposite limit, $|\tau| \gg  1$, we find that $F_P (\tau) \to -
\ln^2|\tau|/(4\tau)$ is  the leading term which  vanishes as~$\tau \to
\infty$.

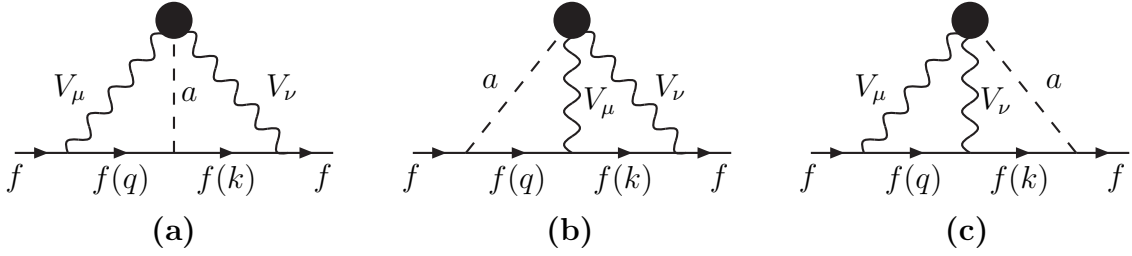
\begin{figure}
 
\begin{center}
\begin{picture}(450,100)(0,0)
\SetWidth{0.8}
 
\ArrowLine(0,30)(20,30)\Text(0,26)[t]{$f$}
\ArrowLine(20,30)(60,30)\Text(40,26)[t]{$f(q)$}
\ArrowLine(60,30)(100,30)\Text(80,26)[t]{$f(k)$}
\ArrowLine(100,30)(120,30)\Text(120,26)[tr]{$f$}

\Photon(20,30)(60,80){3}{5}\Text(27,60)[tr]{$V_\mu$}

\DashLine(60,30)(60,80){5} \Vertex(60,80){7}
\Text(63,55)[tl]{$a$}

\Photon(103,30)(60,80){3}{5.7}\Text(95,60)[tl]{$V_\nu$}

\Text(60,0)[]{\bf (a)}

\ArrowLine(150,30)(170,30)\Text(150,26)[t]{$f$}
\ArrowLine(170,30)(210,30)\Text(190,26)[t]{$f(q)$}
\ArrowLine(210,30)(250,30)\Text(230,26)[t]{$f(k)$}
\ArrowLine(250,30)(270,30)\Text(270,26)[tr]{$f$}

\DashLine(170,30)(210,80){5}\Text(183,60)[tr]{$a$}

\Photon(210,30)(210,80){3}{4} \Vertex(210,80){7}
\Text(215,55)[tl]{$V_\mu$}

\Photon(253,30)(210,80){3}{5.7}\Text(242,60)[tl]{$V_\nu$}

\Text(210,0)[]{\bf (b)}

\ArrowLine(300,30)(320,30)\Text(300,26)[t]{$f$}
\ArrowLine(320,30)(360,30)\Text(340,26)[t]{$f(q)$}
\ArrowLine(360,30)(400,30)\Text(380,26)[t]{$f(k)$}
\ArrowLine(400,30)(420,30)\Text(420,26)[tr]{$f$}

\Photon(320,30)(360,80){3}{5}\Text(330,60)[tr]{$V_\mu$}

\Photon(360,30)(360,80){3}{4} \Vertex(360,80){7}
\Text(365,55)[tl]{$V_\nu$}

\DashLine(400,30)(360,80){5}\Text(390,60)[tl]{$a$}

\Text(360,0)[]{\bf (c)}

\end{picture}
\end{center}
\caption{\it  Feynman graphs  giving  rise to  anomalous fermion  mass
  generation  at three loops.   Black circles  denote the  operator $a
  F_{\mu\nu}\widetilde{F}^{\mu\nu}$ induced by the chiral anomaly (see
  also Fig.~\ref{fig:chiral}).}\label{fig:3loop}
\end{figure}

Our next step is to  evaluate the anomalously generated radiative mass
of  the fermion~$f$, which  results from  the three  three-loop graphs
shown in~Fig.~\ref{fig:3loop}.   By naive  power counting of  the loop
momenta, we  may convince  ourselves that each  of the  three diagrams
(a),  (b)  and  (c)   in  Fig.~\ref{fig:3loop}  possesses  an  overall
logarithmic UV divergence. Specifically, the overall degree $D_\Gamma$
of superficial divergences is found to be
\begin{equation}
  \label{eq:DGamma}
D_\Gamma\ =\ 4 L\: -\: 2 N_{\rm b}\: -\: N_{\rm f}\ =\ 0\; ,
\end{equation}
for  $L=3$ loops,  $N_{\rm b}=3$  boson propagators  due  to $V_\mu$-,
$V_\nu$-  and  $a$-boson  exchanges,  and  $N_{\rm  f}  =  6$  fermion
propagators (two  $f$-propagators along  the $f$-line {\em  plus} four
$F$-propagators contained  in the  triangle graph, including  a chiral
flip  proportional to  $m_F$).  Instead,  one  can show  in a  similar
fashion that all possible one- and two-loop sub-graphs of the diagrams
(a), (b) and (c) in Fig.~\ref{fig:3loop} are UV finite. However, since
$D_\Gamma  = 0$,  one  expects for  the  effective fermion  mass~$m_f$
generated at  three loops to diverge logarithmically  with the cut-off
scale $\Lambda$ in this simple model.

In addition, it is important to observe that the three graphs (a), (b)
and   (c)  of   Fig.~\ref{fig:3loop}  when   added  together   form  a
gauge-invariant  set.   In  order  to demonstrate  this  property,  we
consider   the  gauge-fixing  scheme   of  non-covariant   gauges  for
quantizing the  two gauge-boson propagators in  the loop.  Explicitly,
for the $V_\mu \to V_\alpha$ propagator, we have
\begin{equation}
  \label{eq:Prop}
\Delta_{\mu\alpha} (q)\ =\ \frac{i}{q^2 + i\epsilon}\, \bigg(\!
-\eta_{\mu\alpha}\ + \ \frac{q_\mu \eta_\alpha\: +\: q_\alpha
  \eta_\mu}{q\!\cdot\!\eta}\, \bigg)\; ,
\end{equation}
and a similar expression holds for the $V_\nu \to V_\beta$ propagator.
Here, our convention for the  Minkowski metric is $\eta_{\mu \alpha} =
{\rm  diag}  (1,-1,-1,-1)$,  $\eta_\mu$  is  a  constant  gauge-fixing
null-vector, i.e.~$\eta^2  = 0$,  $V_\alpha$ and $V_\beta$  denote the
gauge  fields attached  along  the $f$-propagator  line  in the  three
graphs (a), (b) and (c)  of Fig.~\ref{fig:3loop}.  It can now be shown
that the  individual gauge-fixing dependence of the  three graphs (a),
(b) and  (c) on  $\eta_{\mu,\nu}$ cancels out  in the sum.   First, we
notice that  all propagator terms  proportional to $q_\mu\eta_\alpha$,
$k_\nu\eta_\beta$, and  $\eta_\mu \eta_\nu$ vanish  identically, after
being  contracted with  the  factor $\varepsilon_{\mu\nu\lambda\rho}\,
k^\lambda q^\rho$  of the $a$-$V^\mu$-$V^\nu$ coupling.   Then, it can
be  shown that  the gauge-dependent  term $q_\alpha  \eta_\mu$  of the
$V_\mu$-propagator in  graph~(a) cancels against  a corresponding term
of the  $V_\mu$-propagator in graph~(b), and  the gauge-dependent term
$k_\beta\eta_\nu$  of~$V_\nu$ in graph~(a)  cancels against  a similar
term  in the  $V_\nu$-propagator in  graph~(c).  Finally,  there  is a
residual gauge  dependence in graphs  (b) and (c) that  still remains,
but this vanishes in their sum.

In  order  to reliably  determine  the  leading  contributions to  the
three-loop  graphs in~Fig.~\ref{fig:3loop},  we consider  an effective
theory  with   a  UV  cut-off  energy  scale~$\Lambda   \le  m_F$  and
approximate     the    loop    function     $F_P(p^2/4m^2_F)$    given
in~(\ref{eq:aVV}) by $F_P(0)=1$, for loop momenta less than $m_F$ (for
a related  effective field-theory approach,  see \cite{Dedes:2012me}).
The  result obtained  for $\Lambda  \ll m_F$  is then  matched  to the
logarithmic  UV  dependence  for   loop  momenta  $\Lambda  \gg  m_F$.
Moreover, we  assume that $m_F$ is  the highest scale  in the problem,
i.e.~$m_F \gg M_a,\ M_V$, which allows us to neglect all other masses.
Since the three graphs  depicted in~Fig.~\ref{fig:3loop} show the same
analytical dependence on  loop momenta less than $\Lambda  = m_F$, the
effective fermion mass $m_f$ may be conveniently expressed as
\begin{equation}
  \label{eq:mfloop}
m_f \ =\ i\, \frac{3\,h_F\,h_f}{4\pi^2}\,
\bigg(\frac{\alpha_V}{4\pi}\bigg)^2\; \frac{Q^2_F\, Q^2_f}{m_F}\;
\gamma_5\, \varepsilon_{\mu\nu\lambda\rho}\, \int\! \frac{d^4k}{\pi^2\,i}\; 
\int\! \frac{d^4q}{\pi^2 i}\ \frac{q^\lambda k^\rho\, \gamma^\nu\!\not\!k
\not\!q\, \gamma^\mu}{(k^2)^2\; (q^2)^2\; (k-q)^2}\ .
\end{equation}
In  order  to  calculate  the  above two-loop  integral,  we  use  the
reduction formula:
\begin{equation}
  \label{eq:redloop}
\int\! \frac{d^4q}{\pi^2 i}\ \frac{q_\alpha q_\beta}{(q^2)^2\;
  (k-q)^2}\ =\ \frac{\eta_{\alpha\beta}}{4}\ \int\! \frac{d^4q}{\pi^2
  i}\ \frac{1}{q^2\; (k-q)^2}\ +\ \frac{1}{2}\; \frac{k_\alpha k_\beta}{k^2}\ .
\end{equation}
Note  that   the  second  term  proportional   to  $k_\alpha  k_\beta$
in~(\ref{eq:redloop}) vanishes when  appropriately contracted with the
anti-symmetric  expression $\varepsilon_{\mu\nu\lambda\rho} k^\lambda$
in~(\ref{eq:mfloop}). Hence,  the reduction formula~(\ref{eq:redloop})
can be  applied twice, first  for the loop  momentum $q$ and  then for
$k$.      In    this     way,    the     effective     fermion    mass
expression~(\ref{eq:mfloop}) simplifies to:
\begin{equation}
  \label{eq:mfsimple}
m_f \ =\  \frac{3\,h_F\,h_f}{2\pi^2}\,
\bigg(\frac{\alpha_V}{4\pi}\bigg)^2\; \frac{Q^2_F\, Q^2_f}{m_F}\;
\int\! \frac{d^4k}{\pi^2\,i}\; \int\! \frac{d^4q}{\pi^2 i}\ 
\frac{1}{k^2\; q^2\; (k-q)^2}\ ,
\end{equation}
where  we   also  used  the  fact  that   $\gamma_5  =  \frac{i}{4!}\,
\varepsilon_{\mu\nu\lambda\rho}\, \gamma^\mu \gamma^\nu \gamma^\lambda
\gamma^\rho$.   By   means  of  a  standard  Wick   rotation,  we  can
analytically     calculate    the    two-loop     integral    occurring
in~(\ref{eq:mfsimple}).   Using spherical coordinates  to parameterize
the  volume   element  of  a  four-dimensional   Euclidean  sphere  of
radius~$\Lambda$, we obtain
\begin{equation}
  \label{eq:IntL}
I_\Lambda \ =\ \int\! \frac{d^4k}{\pi^2\,i}\; \int\! \frac{d^4q}{\pi^2 i}\ 
\frac{1}{k^2\; q^2\; (k-q)^2}\ =\ -\,(4 \ln 2)\; \Lambda^2\; .
\end{equation}
The integral  $I_\Lambda$ is  infra-red safe and  scales quadratically
with    $\Lambda$,   as    expected   by    naive    power   counting.
Substituting~(\ref{eq:IntL}) into (\ref{eq:mfsimple}),  we arrive at a
simple formula for the effective fermion mass of $f$,
\begin{equation}
 \label{eq:L1}
m_f\ =\ -\,6\ln 2\; \frac{h_F\,h_f}{\pi^2}\,
\bigg(\frac{\alpha_V}{4\pi}\bigg)^2\, Q^2_F\, Q^2_f\; 
\bigg(\!\frac{\Lambda^2}{m_F}\!\bigg)\; ,
\end{equation}
which is a good approximation, as long as $\Lambda \ll m_F$.  Then, we
match the  quadratic dependence of  $m_f$ on $\Lambda^2$  obtained for
$\Lambda \ll  m_F$ in~(\ref{eq:L1})  to a logarithmic  dependence $\ln
\Lambda$ for  $\Lambda \gg m_F$.  This  can be done  by observing that
$\ln[(\Lambda^2 + m^2_F)/m^2_F] \to \Lambda^2/m^2_F$, for $\Lambda \ll
m_F$.  Thus, substituting $\Lambda^2/m_F$ with $m_F\, \ln[(\Lambda^2 +
  m^2_F)/m^2_F]$  in~(\ref{eq:L1}),  we   obtain  a  general  estimate
for~$m_f$,
\begin{equation}
  \label{eq:mf}
m_f\ =\ -\,6\,\ln 2\; \frac{h_F\,h_f}{\pi^2}\,
\bigg(\frac{\alpha_V}{4\pi}\bigg)^2\, Q^2_F\, Q^2_f\; m_F\; 
\ln\bigg(\!\frac{\Lambda^2 + m^2_F}{m^2_F}\!\bigg)\; ,
\end{equation}
which  is valid  for  any value  of  the UV  cut-off scale  $\Lambda$.
Assuming  now a  UV cut-off  $\Lambda  = M_{\rm  Pl}$, an  ultra-heavy
fermion~$F$, with mass $m_F =  10^{13}$~TeV close to the GUT scale, an
SO(10)-type  gauge   coupling  $g_V$,  with  $\alpha_V   =  1/30$  and
perturbative  Yukawa  couplings  $h_{F,f}   =  10^{-2}$,  one  gets  a
three-loop effective  fermion mass $m_f \approx 2\times  10^{-9} m_F =
2\times 10^4$~TeV, for $Q_F = Q_f = 1$.  If the pseudoscalar state $a$
has  a mass  $M_a \stackrel{>}{{}_\sim}  m_F$, then  there will  be an
extra suppression factor of $m^2_F/M^2_a$ in~(\ref{eq:mf}), viz.
\begin{equation}
  \label{eq:mfestimate}
m_f\ \sim\ \frac{h_F\,h_f}{\pi^2}\,
\bigg(\frac{\alpha_V}{4\pi}\bigg)^2\, Q^2_F\, Q^2_f\;
\frac{m^3_F}{M^2_a}\; \ln\bigg(\!\frac{\Lambda^2}{M^2_a}\!\bigg)\; .
\end{equation}
Thus, a mild hierarchy of about two orders of magnitude, e.g.~$m_F/M_a
\sim 0.01$, can lead to singlet fermion masses $m_f$ at the TeV scale.

It  is  important to  explore  whether  the  above effective  U(1)$_V$
scenario can be made UV  complete.  A minimal extension is to consider
two pseudoscalars  $a_f$ and $a_F$,  which couple to fermions  $f$ and
$F$, respectively.   Specifically, the Yukawa  sector of the  model is
extended, so as to assume the form
\begin{equation}
  \label{eq:LYa}
{\cal L}_Y\ =\ h_f\, i a_f\, \Big(\bar{f}_L f_R - \bar{f}_R
f_L\Big)\: +\:
h_F\,i a_F\, \Big(\bar{F}_L F_R - \bar{F}_R F_L\Big)\: -\: 
m_F\, \Big(\bar{F}_L F_R + \bar{F}_R F_L\Big)\; .
\end{equation}
The Yukawa Lagrangian~${\cal L}_Y$ possesses two exact symmetries: (i)
CP invariance and (ii)~an $f$-chiral discrete symmetry, where $a_f \to
- a_f$  and  $f_L \to  -  f_L$, whilst  the  fields  $a_F$, $f_R$  and
$F_{L,R}$ do not transform. It is easy to promote these two symmetries
to the  gauge kinetic part  of the Lagrangian. Instead,  we explicitly
break the $f$-chiral discrete  symmetry in the scalar potential~$V$ by
a mixing-mass operator $a_f a_F$ of dimension-2, i.e.
\begin{equation}
  \label{eq:Vpota2}
V (a_f,\ a_F) \ =\ \frac{1}{2}\, M^2_a\, \Big( a^2_f\: +\: a^2_F
\Big)\ +\ \delta M^2_a\, a_f a_F\; .
\end{equation}
For simplicity,  we have taken  here the bilinear mass  parameters for
$a^2_f$ and  $a^2_F$ equal  to $M^2_a$, with  $M^2_a >  \delta M^2_a$.
Consequently,  the  higher   dimension-3  fermion-mass  operator  $m_f
\bar{f} f$ that  breaks the second symmetry~(ii) will  be generated at
the three loop level and it will be UV finite, as we explain below.

The only channel of communication between the $F$- and $f$- sectors is
through the  mixing mass term $\delta  M^2_a\, a_f a_F$  in the scalar
potential~(\ref{eq:Vpota2}).   A  radiative  three-loop mass  for  the
fermion~$f$  will  be  generated   by  the  Feynman  graphs  shown  in
Fig.~\ref{fig:3loop},  where the  $a$-loop exchange  line needs  to be
replaced with  the pseudoscalar transition $a_F \to  a_f$.  Because of
the  latter modification,  the  overall degree  of  divergence of  the
three-loop  graphs reduces  to~$-2$, leading  to a  UV  finite result.
Assuming that  $m_F \gg M_a$,  we may estimate the  three-loop fermion
mass of $f$ to be
\begin{equation}
  \label{eq:mfa2}
m_f\ \sim\  \frac{h_F\,h_f}{\pi^2}\,
\bigg(\frac{\alpha_V}{4\pi}\bigg)^2\, Q^2_F\, Q^2_f\;  
\frac{\delta M^2_a}{M^2_a}\; m_F\; .
\end{equation}
Evidently, if we choose as  before the mass of the ultra-heavy fermion
to be $m_F  = 10^{13}$~TeV, but $h_{F,f} = 1$,  we then obtain fermion
mass $m_f \sim 1$~TeV, for $\delta M^2_a /M^2_a = 10^{-8}$.

It is interesting to discuss the  key differences of a U(1) model with
a CP-even singlet scalar~$\sigma$, instead of a CP-odd boson $a$. Like
the pseudoscalar~$a$, a CP-even  singlet scalar $\sigma$ may also give
rise  to  anomalous  fermion  mass generation,  through  the  operator
$\sigma  F_{\mu\nu} F^{\mu\nu}$  that breaks  anomalously  the scaling
symmetry~\cite{Ellis:1975ap,Shifman:1978zn}.   In this  case, however,
the presence of the explicit scale-violating mass term $m_F \bar{F} F$
of dimension-3 requires  that other explicit scale-violating operators
up to  the same dimension be added  to Lagrangian~(\ref{eq:LU1}), such
as  the  tadpole  $\sigma  T_\sigma$, the  bilinear  mass  $M^2_\sigma
\sigma^2$ and  the trilinear term $\mu_\sigma \sigma^3$,  in order for
the  theory to  remain  renormalizable.  One~may  consider tuning  the
tadpole parameter $T_\sigma$, such that the singlet field $\sigma$ has
a zero  vacuum expectation value~(VEV)  to all orders  in perturbation
theory.   Such  a renormalization  condition  would  prevent $f$  from
acquiring a  mass, through the usual Higgs  mechanism.  However, there
will still  be a two-loop contribution~\cite{Nieves:1981tv,Zee:1985id}
induced by the trilinear term $\mu_\sigma \sigma^3$ that gives rise to
a  UV-divergent  local  mass  for  the  fermion  $f$.   Moreover,  the
trilinear term  $\mu_\sigma \sigma^3$ requires  renormalization beyond
the   tree-level,  as  it   receives  UV-divergent   corrections  from
$F$-fermion  one-loop graphs.   Notice that  in a  CP-invariant theory
containing CP-odd  scalars~$a_{f,F}$ only, model  parameters analogous
to the tadpole $T_\sigma$  and the trilinear coupling $\mu_\sigma$ are
absent to  all orders in  perturbation theory~\cite{Pilaftsis:1998pe}.
For this reason, the observed  CP violation in the quark (and possibly
lepton)  sector  can only  be  explained  within  models that  realize
spontaneous  breaking  of  CP  symmetry,  e.g.~in  two  Higgs  doublet
models~\cite{Lee:1973iz}       (for       a       recent       review,
see~\cite{Branco:2011iw}).   An  additional  requirement is  that  the
ground  states $a_{f,F}$  have zero  VEVs, after  spontaneous symmetry
breaking.

The  three-loop  mass-generating  mechanism  may  have  a  potentially
interesting application  in a  U(1)$_{B-L}$ extension of  the Standard
Model~(SM).  To be specific, we  extend the particle content of the SM
by the  addition of  5 singlet Weyl  neutrinos per  lepton family~$l$:
$n_{1L,R}$, $n_{2L,R}$ and the right-handed neutrino $\nu_R$.  All the
singlet    neutrinos    carry   lepton    number.     In   the    weak
basis~$[(\nu_L)^C\,,\,    \nu_R\,,\,    (n_{1L})^C\,,\,    n_{1R}\,,\,
  (n_{2L})^C\,,\, n_{2R}]$,  the neutrino  mass matrix $M^\nu$  may be
cast into the form:
\begin{equation}
  \label{eq:Mnu6}
M^\nu\ =\ \left(\!\begin{array}{cccccc}
0 & m_D & 0 & 0 & 0 & m'_D \\
m_D & 0 & M^{(1)} & 0 & M^{(2)} & 0 \\
0 & M^{(1)} & 0 & M_1 & 0 &  M^{(3)} \\
0 & 0 & M_1 & 0 & M_2 & 0 \\
0 & M^{(2)} & 0 & M_2 & 0 & M^{(4)} \\
m'_D & 0 & M^{(3)} & 0 & M^{(4)} & M_{B-L} \end{array}\!\right) .
\end{equation}
Here, $M_{1,2}$ are  GUT-scale mass parameters that occur  at the tree
level,  whilst the  parameters $M^{(1,2,3,4)}$  are generated  via the
three-loop  mechanism  discussed  above  and  can be  as  low  as~TeV,
according to  the estimate given in~(\ref{eq:mfa2}).   In addition, we
considered the possibility of  spontaneous breaking of U(1)$_{B-L}$ by
the right-handed  neutrinos $n_2$, where the scale  $M_{B-L}$ could be
close to GUT scale as well. The structure of the mass matrix $M^{\nu}$
may enforced  by demanding that  the fermion fields  $\nu_L$, $\nu_R$,
$n_{2R}$  and  the  pseudoscalar   $a_f$  are  odd  under  a  discrete
transformation, whereas all other neutrino states $n_{1L,R}$, $n_{2L}$
and the pseudoscalar state $a_F$  are even.  Integrating out the heavy
states  $n_{1,2}$ and  assuming that  $m'_D  \ll m_D$,  we obtain  the
low-energy effective neutrino mass matrix
\begin{eqnarray}
  \label{eq:Mnu3}
M^\nu_{\rm eff}\ =\ \left(\!\begin{array}{ccc}
0 & m_D & 0 \\
m_D & 0 & M^{(1)} \\
0 & M^{(1)} & \mu_{B-L} \end{array}\!\right) ,
\end{eqnarray}
where   $\mu_{B-L}   \sim   (M^{(3)})^2/M_{B-L}$   is   an   effective
($B$--$L$)-violating  mass  term.   The  resulting  low-energy  theory
strongly resembles the inverse seesaw model with pseudo-Dirac TeV-mass
heavy  neutrinos~\cite{Mohapatra:1986bd,Nandi:1985uh},  where all  the
isosinglet  masses  are  induced  radiatively.  Likewise,  a  TeV-mass
pattern could also  be obtained for theories with  vector quarks.  The
discussion goes along similar lines and we do not repeat it here.

It  is now  interesting  to observe  that  even in  the  absence of  a
U(1)$_V$ gauge  group, quantum gravity interactions  may be sufficient
to  mediate   the  chirality-violating  effects   of  the  ultra-heavy
fermion~$F$ to the low-energy sector of the theory.  Specifically, one
may envisage a scenario where the  role of the gauge bosons $V_\mu$ is
played by the gravitons  $h_{\mu\nu}$ within a linearized framework of
quantum  gravity.  Then,  integrating  out the  heavy  fermion $F$  in
graphs  analogous to Fig.~\ref{fig:chiral},  an effective  coupling of
the pseudoscalar field  $a$ to gravity will be  generated, through the
operator              $a\,             \varepsilon^{\mu\nu\lambda\rho}
R_{\alpha\beta\mu\nu}\,R^{\alpha\beta}_{\  \   \lambda  \rho}$,  where
$R_{\alpha\beta\mu\nu}$  is  the  Riemann  tensor. We  note  that  the
suggested   mechanism   is  perturbative   pertaining   to  a   matter
pseudoscalar $a$ and it is  not necessarily related to the form-valued
Kalb-Ramond
axions~\cite{Kalb:1974yc,Bowick:1988xh}~\footnote[2]{Recently,      the
  possibility that  Kalb-Ramond axions could still  be responsible for
  the generation of  the operator $a\, \varepsilon^{\mu\nu\lambda\rho}
  R_{\alpha\beta\mu\nu}\,R^{\alpha\beta}_{\   \  \lambda   \rho}$  was
  discussed in~\cite{MP},  within the  context of theories  of quantum
  gravity  with  torsion.}.   Assuming  an  effective  cut-off  energy
scale~$\Lambda = m_F$, we may  perform a naive dimensional analysis of
the respective  three-loop graphs in  Fig.~\ref{fig:3loop} to estimate
$m_f$ as
\begin{equation}
  \label{eq:3loopgrav}
m_f \ \sim\ h_f h_F\; \frac{m^9_F}{m^8_{\rm Pl}}\ \approx\ 10^{-3}\times h_f h_F\; 
\bigg(\frac{m_F}{10^{16}~{\rm GeV}}\bigg)^9\; {\rm GeV}\; ,
\end{equation}
where   $m_{\rm  Pl}  =   M_{\rm  Pl}/\sqrt{8\pi}   \approx  2.4\times
10^{18}$~GeV     is     the      reduced     Planck     mass.      The
estimate~(\ref{eq:3loopgrav})  is  obtained   by  observing  that  the
strength of each gravitational coupling in the loop is proportional to
$1/m^2_{\rm  Pl}$.   If  $h_F  =  h_f  \sim  1$  and  $m_F  =  4\times
10^{16}$~GeV, we find an  effective radiative mass $m_f \sim 250$~GeV,
induced  by ordinary  quantum gravity  effects.  This  result  is very
sensitive to  the mass $m_F$ and  the actual UV  completion of quantum
gravity.   Nevertheless,  it  is  amusing  to note  that  for  $m_F  =
10^{16}$~GeV, one may  account for isosinglet masses $m_f$  in the keV
range. In particular,  it was argued~\cite{Asaka:2006ek} that keV-mass
sterile neutrinos may successfully play the role of warm dark matter.

It would  be worth investigating whether  a supersymmetric realization
of  our  anomalous  three-loop  mechanism exists.   Let  us  therefore
consider  an anomalous-free  U(1)$_V$  model with  one chiral  singlet
superfield $\widehat{S}$  and two  pairs of oppositely  charged chiral
superfields    $\widehat{F}_{L,R}$    and~$\widehat{f}_{L,R}$.     The
pertinent superpotential of this model is given by
\begin{equation}
  \label{eq:Wpot}
W\ =\ h_F\, \widehat{S}\,\Big( \widehat{F}_L \widehat{F}_R\, -\, 
m_F^2 \Big)\: +\: M_S\,\widehat{S}^2\: +\: h_f\, \widehat{S}
\widehat{f}_L \widehat{f}_R\; .
\end{equation}
In the  absence of the  term $m^2_F$, the superpotential  possesses an
exact  $R$   symmetry,  with   $R$  charges:  $R(\widehat{S})   =  1$,
$R(\widehat{F}_{L,R}) =  R(\widehat{f}_{L,R}) = 1/2$ and $R  (W) = 2$.
The  superpotential tadpole  parameter $m^2_F$  breaks softly  the $R$
symmetry  and may  effectively  be generated  by supergravity  effects
(see,  e.g.~\cite{Bagger:1995ay} and  references  therein).  For  this
reason,  we assume  that  both  $m_F$ and  $M_S$  are large  GUT-scale
parameters. On  the other  hand, according to  the non-renormalization
theorem   in   supersymmetry,   a   superpotential  mass   term   $m_f
\widehat{f}_L   \widehat{f}_R$    cannot   be   generated    by   {\em
  re\-normalizable}  supersymmetric  interactions  to  all  orders  in
perturbation.  Also, the existence of  such a term would break the $R$
symmetry softly by a dimension-3 operator.

In the supersymmetric limit, one possible solution to the ground state
of  the above  supersymmetric  model consists  of  having the  singlet
supefield    $\widehat{S}$   with    vanishing   VEV~\footnote[3]{Soft
  supersymmetry breaking effects may  generate a non-zero VEV $\langle
  S \rangle \sim  (A_F - \xi_S)/h_F$, which can  be naturally small of
  the  electroweak order,  or  even  tuned to  zero,  where $A_F$  and
  $\xi_S$  are the soft  supersymmetry breaking  parameters associated
  with    the    trilinear    coupling   $\widehat{S}    \widehat{F}_L
  \widehat{F}_R$   and  the   tadpole   $\widehat{S}$.},  whilst   the
U(1)$_V$-charged  chiral  superfields  $\widehat{F}_{L,R}$  develop  a
non-zero    VEV,   i.e.~$\langle   \widehat{F}_L\rangle    =   \langle
\widehat{F}_R  \rangle = m_F$,  thus breaking  U(1)$_V$ spontaneously.
The   other  charged   chiral  superfields   $\widehat{f}_{L,R}$  have
vanishing VEVs and so their  fermion components remain massless in the
supersymmetric  limit.  Instead, the  corresponding fermion  fields of
$\widehat{F}_{L,R}$   combine   with   the  fermionic   component   of
$\widehat{S}$  and  the  gauginos  of  U(1)$_V$ to  form  heavy  Dirac
fermions of  masses of order  $m_F$ and $M_S$.  Integrating  out these
heavy  fermions,  we obtain  the  5-dimensional operator  $\widehat{S}
\widehat{W}^\alpha \widehat{W}_\alpha$,  where $\widehat{W}_\alpha$ is
the chiral  superfield strength spinor  related to the  U(1)$_V$ gauge
group. Even though this operator  is analogous to the one generated in
Fig.~\ref{fig:chiral},    the     non-renormalization    theorem    in
supersymmetry prohibits  the generation of  a three-loop mass  by loop
graphs  similar  to   Fig.~\ref{fig:3loop}  and  their  supersymmetric
counterparts.   However, non-renormalizable  supergravity interactions
might be  sufficient to generate  a sizeable effective  radiative mass
for   $m_f$,   of   the    electroweak   order,   according   to   the
estimate~(\ref{eq:3loopgrav}),  when  the  mass parameters  $m_F$  and
$M_S$ are taken to be  sufficiently large, e.g.~in the vicinity of the
gauge coupling  unification scale $M_X \sim  2\times 10^{16}$~GeV.  An
extensive calculation lies beyond the scope of this note and may given
elsewhere.

In  summary,  we  have  presented  a  novel  radiative  mechanism  for
generating  non-zero fermion  masses through  global anomalies  at the
three  loop  level.   A   minimal,  UV-complete  realization  of  this
mechanism requires  the presence of at least  two singlet pseudoscalar
states which  could couple  to gauge-invariant fermion  bilinears.  We
have assumed that one family~$F$ of fermion bilinears breaks chirality
at a very high scale, e.g.~close to the GUT scale.  In a gauge theory,
this  chirality violation  can be  mediated to  another family  $f$ of
fermions  by  the  three-loop  graphs  shown  in~Fig.~\ref{fig:3loop},
producing masses that could be  of the electroweak order, according to
our  estimate  in (\ref{eq:mfa2}).   This  mechanism  can  be used  to
naturally  explain the possible  existence of  heavy neutrinos  at the
electroweak or even  lower scale.  In addition, we  have discussed its
potential implications  for model building,  including the possibility
of  creating low-scale fermion  masses by  quantum gravity  effects in
non-supersymmetric and  supersymmetric theories.  Further  studies are
needed in this direction, in order to be able to assess the full range
of applicability of the proposed mechanism.

\bigskip

\subsection*{Acknowledgements} This work is supported in part by the
Lancaster--Manchester--Sheffield  Consortium for  Fundamental Physics,
under  STFC research grant:  ST/J000418/1. The  author also  wishes to
thank the  Galileo Galilei Institute  for Theoretical Physics  for the
hospitality and the INFN for  partial support during the completion of
this work.

\end{document}